\documentclass[conference,10pt]{IEEEtran}
\IEEEoverridecommandlockouts

\usepackage{cite}
\usepackage{amsmath,amssymb,amsfonts}
\usepackage{graphicx}
\usepackage{textcomp}
\usepackage{xcolor}
\let\labelindent\relax
\usepackage{enumitem}
\usepackage{romanbar}
\usepackage{amsmath}
\usepackage{amsmath}
\usepackage{tikz}
\usepackage{mathdots}
\usepackage{yhmath}
\usepackage{cancel}
\usepackage{color}
\usepackage{siunitx}
\usepackage{array}
\usepackage{multirow}
\usepackage{amssymb}
\usepackage{gensymb}
\usepackage{tabularx}
\usepackage{booktabs}
\usepackage{verbatim}
\usepackage{caption}

\usepackage[utf8]{inputenc} 
\usepackage{textgreek}
\usepackage{algorithm}
\usepackage{algpseudocode}

\usepackage{caption}
\usepackage{subcaption}

\usetikzlibrary{fadings}
\def\BibTeX{{\rm B\kern-.05em{\sc i\kern-.025em b}\kern-.08em
		T\kern-.1667em\lower.7ex\hbox{E}\kern-.125emX}}
\begin{document}
	
 		\title{5G D2D Transmission Mode Selection Performance \& Cluster Limits Evaluation of Distributed Artificial Intelligence and Machine Learning Techniques		
		\thanks{This research is part of a project that has received funding from the European Union's Horizon 2020 research and innovation programme under grant agreement Nº739578 and the government of the Republic of Cyprus through the Directorate General for European Programmes, Coordination and Development.}
	}
	
	\author{\IEEEauthorblockN{Iacovos Ioannou\authorrefmark{1}\authorrefmark{2}, Christophoros Christophorou\authorrefmark{1}\authorrefmark{3}, Vasos Vassiliou\authorrefmark{1}\authorrefmark{2}, and Andreas Pitsillides\authorrefmark{1}}
		\IEEEauthorblockA{\authorrefmark{1}Department of Computer Science, University of Cyprus\\
			\authorrefmark{2}CYENS Center of Excellence, Nicosia, Cyprus \\
			\authorrefmark{3}CITARD Services LTD, Nicosia Cyprus \\
		}
	}
	
	\maketitle
	
	\begin{abstract}
		5G D2D Communication promises improvements in energy and spectral efficiency, overall system capacity, and higher data rates. However, to achieve optimum results it is important to select wisely the Transmission mode of the D2D Device in order to form clusters in the most fruitful positions in terms of Sum Rate and Power Consumption. Towards this end, this paper investigates the use of Distributed Artificial Intelligence (DAI) and innovative to D2D, Machine Learning (ML) approaches (i.e., DAIS, FuzzyART, DBSCAN and MEC) to achieve satisfactory results in terms of Spectral Efficiency (SE), Power Consumption (PC) and execution time, with the creation of clusters and backhauling links in D2D network under existing Base Station. Additionally, this paper focuses on a small number of Devices (i.e., \textless=200), with the purpose to identify the limits of each approach in terms of low number of devices. More specifically, we investigate when an operator must consider implementing a D2D network (that requires extra complexity), therefore when the cluster members are sufficient enough to achieve better results than the classic mobile network. So, this research identifies where it is beneficial to form a cluster, investigate the critical point that gains increases rapidly and at the end examine the applicability of 5G requirements. Additionally, prior work presented a Distributed Artificial Intelligence (DAI) Solution/Framework in D2D and a DAIS Transmission Mode Selection (TMS) plan was proposed. In this paper DAIS is further examined, improved in terms of thresholds evaluation (i.e., Weighted Data Rate (WDR), Battery Power Level (BPL)), evaluated, and compared with other approaches (AI/ML). The results obtained demonstrate the exceptional performance of DAIS and FuzzyART, compared to all other related approaches in terms of SE, PC, execution time and cluster formation. Also, results show that the investigated AI/ML approaches are also beneficial for Transmission Mode Selection (TMS) in 5G D2D communication, even with a smaller numbers of devices (i.e., \textgreater=5 for clustering, \textgreater=50 for back-hauling) as a lower limits.
	\end{abstract}
	
	
	\section{Introduction}
	\label{intro}
	
	Device-to-Device (D2D) Communication is expected to be a contributing factor in achieving the demanding requirements of 5G Mobile Communication Networks \cite{Akyildiz2016,Ioannou2020}. The main reasons are that D2D communication is not constrained by the licensed frequency bands and that it is transparent to the cellular network. Also, it permits adjacent User Equipment (UE) to bypass the Base Station (BS) and establish direct links between them, either by sharing their connection bandwidth and operate as relay stations, or by directly communicating and exchanging information. For the aforesaid reasons, D2D can improve spectral efficiency, data rates, throughput, energy efficiency, delay, interference and fairness \cite{Ioannou2020,Fodor2012,Gandotra2016,Ahmad2017}.
	
	However, in order to achieve optimum results, it is important, among others, to select wisely the Transmission Mode of the D2D Device in order to form clusters in the most fruitful positions in terms of Sum Rate and Power Consumption. The main reason is that the Transmission Mode selection for a device can affect the creation of the clusters, the way devices will form links and communicate among them, and it can also optimize backhauling links between disconnected/disjointed clusters by forming better paths.

	Towards this end, our previous work \cite{Ioannou2020} proposed: \textbf{i)} a BDIx (BDI extended) agents based Distributed Artificial Intelligence (DAI) Framework that can achieve D2D communication in an efficient and flexible manner by focusing on the local environment rather the global environment with the use of the DAIS Plan. For assessing the efficiency of the DAIS approach, threshold values affecting spectral efficiency and power usage of the network, like the Weighted Data Rate (WDR) and the Battery Power Level (see Section \ref{section:problemFormulation}) of the D2D Device, have been employed. In addition, those achieving strong performance have been determined. The effect of the Transmission Power (TP) variation of each Device on the investigated approaches, in terms of total Spectral Efficiency (SE) and Power Consumption (PC) was also examined. This investigation focuses on D2D communication network with a small number of devices for the following reasons: i) applicability of 5G requirements; ii) investigate the critical point that gains increases rapidly; iii) coverage expansion; and iv) find the limits of the approaches in terms of better SE and PC than non-D2D UE approach (classical mobile approach).	
		
	In this paper, the efficiency of DAIS is further examined, evaluated, and compared in terms of SE, PC, Execution Time (ET), cluster formation characteristics (i.e., number of messages exchange, resulting non-D2D UEs, mean number of devices per cluster, number of cluster created)  with other related approaches. More specifically, we compare DAIS with Distributed Random, Sum Rate Approach, Centralized non-D2D-UE (shown in \cite{Ioannou2020}) and other, currently introduced to D2D and Transmission Mode Selection Artificial Intelligence/Machine Learning (AI/ML) techniques (i.e., FuzzyART \cite{Carpenter1991,Akojwar2006}, DBSCAN \cite{Ester96adensity-based,Li2016} and MEC \cite{Golchin1997,HaifengLi2004}) in a 5G D2D communication network with a reduced number of devices (\textless= 200 UEs/D2D candidates). Note that FuzzyART, DBSCAN and MEC are centralized unsupervised learning clustering techniques that, for the purposes of this research, we utilized for D2D communication. These approaches do not require a learning process in order to be used in the D2D communications and they provide good clustering results. The underlying reasons for selecting unsupervised learning clustering techniques are the following: \textbf{i)} the Transmission Mode Selection is directly associated with the selection of best Cluster Head, therefore the clustering techniques must be used; and \textbf{ii)} due to the dynamic nature of mobile communication network the training part of supervised learning can not conclude to the best results because of the devices movement and due to the fact that in D2D communication the best data are the current data.

	The results obtained demonstrate that with the right tuning of the thresholds, DAIS could provide significant improvement in the network. Furthermore, from the results obtained from the comparison of the investigated approaches it was observed that DAIS along with FuzzyART outperforms all other approaches, except Sum Rate Approach, in terms of total SE and total PC. The reason that Sum Rate Approach achieved better results than DAIS and FuzzyART is because Sum Rate Approach has a global knowledge of the network and thus can select the best transmission mode. Even so, DAIS approaches the performance of the Sum Rate Approach and achieves similar results with FuzzyART, acting on only local information. In addition, it was observed that Transmission Power (TP) alteration of the D2D Devices with a small number of UEs (\textless=200) can affect SE and PC for all investigated approaches. Also, results show that the investigated AI/ML approaches are also beneficial for Transmission Mode Selection (TMS) in 5G D2D communication, even with a smaller numbers of devices (i.e., \textgreater=5 for clustering, \textgreater=50 for back-hauling) as a lower limits..	 	

	The rest of the paper is structured as follows. Section \ref{baground_knowledge_related_work} provides some background information about D2D, the investigated Unsupervised Learning Clustering (ULC) techniques and related work associated with approaches that tackle transmission mode selection. More specifically, it provides a brief description of the DAI Framework and DAIS Plan along with some specifics about the investigated approaches (i.e., DAIS, Sum Rate, Distributed Random, non-D2D UE) that are introduced in the \cite{Ioannou2020}, such as the implementation, assumptions, constraints, thresholds and utilized metrics. Furthermore, this section also provides information about other approaches that tackle Transmission Mode Selection from open literature. Section \ref{problem_formulation_and_approaches_specifics} presents the problem that this paper tackles and provides some additional assumptions of our examination along with some extra information about the formulation. Section \ref{ULheuristic} provides the implementation of a heuristic algorithm that is using the UCL approaches results, in order to achieve Transmission Mode Selection in D2D communication. Additionally, it provides the assumptions, constraints and metrics of the ULC approaches. The efficiency of the investigated approaches, is examined, evaluated and compared in Section \ref{performance_evaluation}. Finally, Section \ref{conclussions_and_future_work} contains our Conclusions and Future Work.
	
	\section{Background Knowledge and Related Work}\label{baground_knowledge_related_work}
	
	\subsection{Background Knowledge}\label{CD2D}
	This section provides background knowledge regarding the main characteristics of D2D communications. More specifically, the types of control that can be exploited for the establishment of D2D communication links, as well as the types of transmission modes that a D2D Device can operate, are outlined in this section. Additionally, in this section a brief description is provided for each of the Unsupervised Learning Clustering approaches that are investigated.
	
	\subsubsection{Types of Control in D2D Communication}
	\label{TypesOfControl}
	The types of control that can be used for the establishment of D2D Communication links can be categorized as follows: i) Centralized: The Base Station (BS) completely oversees the UE nodes even when they are communicating directly; ii) Distributed: The procedure of D2D node management does not oblige to a central entity, but it is performed autonomously by the UEs themselves; iii) Distributed Artificial Intelligence (DAI): All control processes run in parallel and begin at the same time through collaboration in an intelligent manner; and iv) Semi distributed/hybrid: A mix of centralized and distributed schemes.
		
	\subsubsection{Types of Transmission Modes in D2D Communication}
	The different transmission modes in D2D Communication are the following: i) D2D Direct: Two UEs connect to each other by utilizing licensed or unlicensed spectrum; ii) D2D Single-hop Relaying: Contribution of bandwidth between a UE and other UEs \cite{Deng2015a}. One of the D2D UEs is connected to a BS or Access Point and provides access to an additional D2D UE; iii) D2D Multihop Relay: The single-hop mode is extended by empowering the connection of more D2D UEs in chain. This chain can be one to one relationship or one to more\cite{Steri2016}; iv) D2D Cluster \cite{Song2014}: D2D Cluster is a group of UEs (D2D Devices acting as clients) connected to a D2D relay node performing as a Cluster Head (CH)\cite{Peng2013}; and v) D2D Client: D2D Client is the selection of UE to participate in a D2D Cluster and act as client. 

	\subsubsection{Investigated Unsupervised Learning Clustering Approaches}\label{UL}	
	\paragraph{FuzzyART \cite{Carpenter1991,Akojwar2006}} \label{FUZZY_ART}
	Fuzzy ART is an unsupervised learning clustering algorithm. It is a type of Adaptive Resonance Theory (ART) network approach, which similarly to K-Means algorithm uses single prototypes to internally represent and dynamically adjust clusters. However, Fuzzy ART uses a minimum required similarity between patterns that are categorized within one cluster. The resulting number of clusters depends on the distances between all input patterns, presented towards the network for the period of training cycles. Fuzzy ART uses structure calculus based on fuzzy logic and ART for binary and continuous value inputs.
	
	\paragraph{DBSCAN \cite{Ester96adensity-based,Li2016}} \label{DBSCAN}
	The DBSCAN relies on a density-based concept of clusters which is outlined to determine clusters of uninformed shape. The fundamental concept behind DBSCAN is that a point belongs to a cluster if it is close to many points from that cluster.  More specifically, for each point of a cluster, the neighborhood of a given radius has to enclose at least a minimum number of MinPts points (called core point), If there are at least minPts number of points in the neighborhood, the point is marked as core point and a cluster formation starts. If not, the point is marked as noise. In the direction of finding a cluster, it starts with a random point and retrieves all points density-reachable from the chosen point. A border point is reachable from a core point and there are less than minPts number of points within its surrounding area. Additionally, an outlier point is not a core point and not reachable from any core points. In DBSCAN, radius (called Eps) is the radius of the neighborhood and minimum points of a cluster (called MinPts) are the minimum number of points in the G-neighborhood of a core point. The aforesaid parameters are respectively mandatory parameters to the algorithm. Moreover, the recursive clustering of the points of a cluster is only crucial under conditions that can be uncomplicatedly recognized with the use of the Euclidean distance.

	\paragraph{Minimum Entropy Clustering (MEC) \cite{Golchin1997,HaifengLi2004}} \label{MEC}
	Minimum Entropy Clustering (MEC) algorithm, focuses on the minimization of the conditional entropy of clusters, given samples so at the end it concludes with the clusters. Numerous mathematical facts, such as Fano’s inequality and Bayes probability of error, indicate that the MEC method can perform well on grouping patterns. This is the reason that MEC: i) performs well even when the correct number of clusters is unknown; ii) correctly reveals the structure of data; and iii) effectively identifies outliers simultaneously. However, MEC is an iterative algorithm starting with an initial partition given by any other clustering methods (e.g., K-Means) except the random initialization. Therefore, in this investigation the initialization is done with the use of the data results coming from the K-Means\footnote{K-Means (Lloyd's algorithm) forms \textit{K} clusters, based on the total number of samples (e.g., in our case the number of the UEs in the Network), in which each sample belongs to the cluster with the nearest mean.} execution. The method can also correctly reveal the structure of data and effectively identify outliers simultaneously with the minimum entropy clustering criterion.

	\subsection{Related Work}\label{RW}
	This section provides a brief description of the Distributed Artificial Intelligence (DAI) Framework along with its Desire Plan DAIS together with Sum Rate and Distributed Random algorithms that perform Transmission Mode Selection as shown in \cite{Ioannou2020}. Additionally, this section provides information regarding FuzzyART, DBSCAN, and MEC unsupervised learning Machine Learning (ML) clustering techniques, and other related approaches from open literature on Transmission Mode selection in D2D Communication. It is important to highlight here that the aforesaid AI/ML techniques were not designed for application in D2D communication but they are utilized and applied to D2D communication by us, for the purposes of this research, due to their scalability, metric used, parameters and way of calculation of labels of clusters.
		
	\subsubsection{Distributed Artificial Intelligent (DAI) Framework}
	In this section, the paper explains in brief the DAI Framework that as concept it was introduced in the \cite{Ioannou2020}. The main objective of the DAI framework is to implement 5G D2D communication with the purpose to achieve the D2D challenges (as shown in \cite{Ioannou2020}). By enabling D2D UEs through BDIx agents that instantiate through BDIx framework, the investigation aims for the devices to act independently, autonomously and as a self-organizing network. More precisely, in order to achieve the aforementioned characteristics, the framework it utilizes software agents and especially BDI (Belief-Desire-Intention) agents with extended Artificial Intelligence/Machine Learning capabilities (ex. Neural networks, Fuzzy logic) named as BDIx Agent. The framework acts as a glue in the employment of more than one of successful, optimized intelligent technologies (e.g. Neural Networks, Fuzzy Logic). Therefore, the BDIx framework will be modular and the believes and desires can be substituted, added by any proposed approach that will have as target to achieve the D2D communication in 5G, as long the stability of the agent is achieved. Additionally, such agents in the framework can be implemented at the UEs as a software and there is no need to change how BSs operate or to change the hardware at BSs or UEs.

	\paragraph{Distributed Artificial Intelligence Solution (DAIS) Approaches \cite{Ioannou2020}} \label{dais}
	DAIS is Plan implemented in a BDIx agent based DAI Framework, that selects the transmission mode of a D2D Device in a distributed artificial intelligence manner (as shown in the \cite{Ioannou2020}). More specifically, the DAIS is related with several Desires (e.g., Signal quality is acceptable, Data Rate is acceptable, WDR is acceptable). So, when any of these approaches get priority 100\% and become Intentions, is start execution. More specifically, with the network event of "D2D Device entering in D2D communication network" the  "WDR is acceptable" priority gets the priority value of 100\% and executes DAIS. Therefore, for the Transmission mode selection, the WDR (Weighted Data Rate), a new metric that we introduced in \cite{Ioannou2020}, is considered. So, with the implementation of DAIS and the use of BDIx agents, there are some assumptions, constraints, thresholds, and a new metric that are introduced. However, in order to show how the BDIx Agents framework can be optimized in terms of threshold investigation, only the ”Weighted Data Rate” (WDR)\footnote{The WDR metric is defined at each node in D2D communication as the minimum data rate in the path that the UE has selected, either this is directly connected to the BS or through another D2D Device.} metric has been analyzed and utilized. Basically, the aim of the DAIS approach is to maximize its WDR (i.e., WDR = max(min(Link Rate))) by selecting the path with the highest WDR among the surrounding paths around the D2D Device. In our previous research we introduced some thresholds (WDR/BPL Threshold). A brief description of the thresholds is shown below:
	
	\begin{itemize}
		\item The Battery Power Level threshold determines the minimum value (in \%) that a D2D Device must have in the remaining battery, in order to become D2D Relay or D2D Multi-Hop Relay and accept connections from other UEs. More specifically, a D2D Relay or D2D Multi-Hop Relay Device will admit connections from new D2D Devices entering the Network only when their battery power level is greater than or equal to the battery threshold. The reason for utilizing the battery power level threshold is: i) fairness; ii) network stability and iii) longevity.
		\item The WDR threshold determines: i) the minimum WDR that an existing D2D Device operating as D2D Relay/ D2D Multi-Hop Relay must have in order for a new D2D Device entering the network to connect to it; or ii) the maximum WDR that a new D2D Device entering the D2D Network must have in order to replace a D2D Device operating as D2D Relay/ D2D Multi-Hop Relay and take its role. The WDR threshold is used by the algorithm for four purposes. More specifically, through the WDR threshold, new D2D Device entering in the Network:
		\begin{itemize}
			\item Can perform a quality check of the D2D Relay, in order to connect to it as a D2D Client.
			\item Can perform a quality check of the D2D Multi-Hop Relay, in order to connect to it either as a D2D Client or a D2D Relay.
			\item Can perform a replacement of a D2D Relay/ D2D Multi-Hop Relay device and take its role, if the new D2D device's WDR is greater than the WDR of the existing D2D Relay/D2D Multi-Hop Relay device.
			\item Can connect to a D2D Relay/ D2D Multi-Hop Relay Device in its proximity, and act as a D2D Relay.
		\end{itemize}		
	\end{itemize}

	\subsubsection{Sum Rate, Distributed Random and non D2D UE Approaches \cite{Ioannou2020}} \label{PW}
	Sum Rate Approach, is a distributed intelligent approach which uses the sum rate of the network as a metric for the UE Device to select the best Transmission mode. Note that in the Sum Rate Approach the D2D Device selects the most appropriate Transmission Mode by having all the knowledge of the network (i.e., D2D Relays, D2D Multi Hop Relays, D2D Clients, connection links).	More precisely, by having the complete network knowledge (i.e., Data Rate of each link at each device, Device selected Transmission Mode), each node adds the data rate of the connections that each D2D Device has in the D2D communication network. Then it decides the best transmission mode, best link and best path to the BS or other Gateway, in order to achieve the maximum Sum Rate of the whole network. On the other hand, the Distributed Random approach is a distributed approach which performs Transmission mode selection in a random manner (e.g. the algorithm for Transmission Selection selects randomly a mode of the entering device). The non D2D UE approach describes the current approach used in Mobile Networks. This approach keeps all the UEs connected directly to the BS and a constant predefined transmission power, that is specified for the UEs that are directly connected to the BS, is used.	
	
	Overall in our approach we consider as the worst case scenario the Random approach and the best approach as the Sum Rate approach that knows all the D2D Devices and the links in the D2D network with the opportunity to do a brute force calculation, with the target to calculate the maximum possible SE that results to reduced PC.

	\subsubsection{Related work on Transmission Mode Selection in D2D Communication}\label{RTS}
	Approaches related to the Transmission mode selection investigated in this paper, are provided in a plethora of articles \cite{Doppler2010,Jung2012,Xu2017,Ma2012,Rigazzi2014,Gui2018,Ioannou2020}. The metrics considered for selecting the transmission mode to be adopted are: power, interference, resource blocks (RB), SINR, distance, power, frequencies and WDR. In the literature one can find approaches with a focus on: i) D2D Device Selection \cite {Doppler2010,Jung2012,Liu2016}; ii) Relay selection only \cite{Xiang2012,Xu2017,Ma2012}; and iii) D2D multi-hop relay forming by selecting as modes the D2D or D2D Multihop \cite{Rigazzi2014,Gui2018}. In our work we are examining all of the possible transmission modes that can be assign to a UE, by itself (e.g. BDIx Agent) or by other entities (e.g. BS).
	
	A classification on the related approaches based on the type of control (see Section \ref{TypesOfControl}) is: i) Centralized \cite{Doppler2010,Jung2012,Xu2017,Rigazzi2014,Gui2018}, where the decision is taken by the BS; ii) Semi-distributed approaches \cite{Liu2016}, where the decision is taken by both the BS and the D2D Devices in collaboration; iii) Distributed \cite{Ma2012}, where the decision is taken by the D2D Devices; however in this case the D2D Devices need some information from the BS; and iv) Distributed Artificial Intelligent (DAI) \cite{Ioannou2020}, where the decision is taken by each D2D Device independently; however, in this case they may share information with other D2D Devices.
	
	It is evident from the above preliminary survey that most works use the Centralized approach and only a few use Semi or Fully Distributed algorithms. Additionally, we could not identify any other approach in the open literature that tackles the problem of having a D2D Device utilizing all transmission modes (D2D Relay, D2D Multi-Hop Relay and D2D Cluster) in a distributed AI manner. Furthermore, to the best of our knowledge, there is not any other D2D transmission mode selection approach in the literature that is utilizing unsupervised learning AI/ML clustering techniques. Therefore, the usage of unsupervised learning AI/ML approaches for the Transmission mode selection in D2D communication, is also a contribution of this paper.
	
	\section{Problem Description}\label{problem_formulation_and_approaches_specifics}
	\label{section:problemFormulation}
	In this paper we aim to use DAI and ML in order for a D2D Device to select a Transmission Mode and create a D2D communication network with for the purpose to reduce the distance to the Access Point, reduced the latency, increase SE and reduced PC in a small (\textless=200) number of Devices D2D Network. The number of UEs examined is small, due to one of the major contribution of the paper, because we aim to calculate the investigated approaches lower limits in an environment of small number of devices in order to to show where is fruitfully to achieve cluster with drones, other relay devices and an operator should consider not change the topology of the network. Additionally, please note that similar problems, even with the same number of devices, are resolved with the use of small cells \cite{smallcell1},\cite{smallcell2}.

	Additionally, in order for the  Clusters in D2D (though transmission mode selection) to achieve higher Sum Rate (Total Spectral Efficiency) and reduced total Power Consumption values, and increase quality\footnote{By having a large number of D2D Devices under a cluster the total SE is increased, total PC is reduced and the number of direct links to BS are decreased. On the other hand, in the case of a large number of Clusters the links to BS are reduced but SE may not be affected. Balancing of both metrics can be achieved with maximum SE, minimum PC and reduced number of links to BS for large towards medium number of clusters with equal assigned D2D Client Devices.}, there are factors that affects the cluster forming quality by an approach. The major contributing factors in the successful realization of a quality D2D clusters under a D2D network are the following: i) number of messages exchanged in order for an approach to conclude; ii) resulting of non D2D UEs; iii) number of clusters created; iv) total number of devices under a cluster; v) position of Cluster Head (CH); and vi) Data Rate of CH. The factors from i to iv, are examined for all the investigated approaches in order to examine their applicability at transmission model selection in D2D. The rest of factors are not examined due to the direct relation of them to the inner workings of each investigated approach.

	Therefore, the problem that this paper tries to tackle is sixfold:	
	\begin{itemize}
		\item It tries to maximize the total SE (i.e, sum rate) and reduce the total PC of the DAIS algorithm as well as the other investigated unsupervised learning AI/ML clustering techniques, in the case of a small number N of devices (\textless=200 UEs) under a BS. Therefore, this paper have the following assumptions about the physical link:
			\begin{itemize}
		        \item The D2D network consists of N devices under the Base Station (BS)
		        \item Our approach focuses on the mobile and wireless networks with a single-antenna and point-to-point scenario
		        \item Our approach uses the Free Space Model and Free Space Path Loss
		        \item Our approach uses the Additive White Gaussian Noise (AWGN) as the basic noise model
		        \item The Transmission Power (TP) is known 
		        \item The Spectral Efficiency is calculated per link
		    \end{itemize}
		\item It examines the problem of forming Back-hauling links, with the selection of D2D Multi Hop Relay Transmission mode, form DAIS and Sum Rate in small number of devices network. 
		\item It examines the problem of identifying the best cluster heads in a D2D communication network with the use of Transmission Mode Selection and AI/ML techniques for the Unsupervised Learning Clustering techniques.
		\item It examines if unsupervised learning techniques can be utilized in order to achieve equal or better results than DAIS and/or Sum Rate Approach, in terms of Transmission mode selection (as shown at \cite{Ioannou2020}).
		\item It examines the time that each approach used for structuring the D2D communication network.
		\item It examines the limits, in terms of when is valuable for the clustering algorithms to be used in a Mobile environment according to number of UEs under the BS. Additionally, we examine when for the approach is valuable to implement Back-hauling and use D2D Multi Hop Relay Transmission Mode.
	\end{itemize}

	In this paper an investigation of the DAIS thresholds is executed with the purpose to increase the Total SE and Total PC (the formulation of SE and PC and problem formulation is shown at \cite{large}).

	\section{Unsupervised Learning Clustering Approaches Utilization in D2D communication with a Heuristic Algorithm}\label{ULheuristic}
	A heuristic algorithm (see section \ref{performance_evaluation}) has been developed that utilizes the clustering results extracted by FuzzyART, DBSCAN and MEC approaches to select the best D2D Device in the identified cluster to be set as a D2D Relay node. Note that the metric used to perform the selection is the Data Rate (as described in Algorithm \ref{alg}). Likewise, the feature set used for all the unsupervised learning clustering approaches is the same and it is the set composed with latitude and longitude (coordinate). Additionally, note that the aforementioned approaches does not form backhauling more than one hop and the selection of D2D Multi Hop Relay is not provided as  selection option of Transmission Mode in the approaches. Therefore, the Transmission Mode that Approach Devices can select is D2D Relay and D2D Client.   
	
	It is worth mentioning that in order to apply the FuzzyART, DBSCAN and MEC approaches to the needs of D2D Communication, we utilized these approaches and set the constraints/settings set out below:
	\begin{itemize}
		\item For all approaches, we set the maximum radius distance to form a cluster to 200 meters (WiFi Direct).
		\item For FuzzyART we do not limit the maximum number of clusters allowed (maxClusterCount=-1).
		\item For DBSCAN we set the minimum points (minPts) of the cluster to 2.
		\item For MEC we set the number of clusters (k) to 100 (note that the final number of clusters may be less).		
	\end{itemize}	
	
	Note that except from the aforesaid constraints/settings set, all other default settings and constraints provided by the “SMILE” framework are the same \cite{SMILE}.
	
	\begin{algorithm}
		\begin{algorithmic}[1]
			\fontsize{7}{9}\selectfont	
			\State i: radius of Cluster Head
			\State T: a set containing clusters		
			\Procedure{ClusterHeadDetection}{$T_{th},i$}
			\State $T_{u_{i}} \gets list\ of\ Clusters from T_{th}$
			\For{each cluster $c$ in $T_{u_{i}}$}
			\State $Node{c_i} \gets maximum Data Rate in cluster c$
			\State $Nodes{c_i} \gets list\ of\ Nodes\ from\ c$
			\For{each node $n$ in $Nodes{c_i}$}
			\State \sc{We have two dimensions of each coordinate (latitude,longitude) for euclidean distance}
			\State $d\left( n,Node{c_i}\right) = \sqrt {\sum _{j=1}^{2} \left( n_{j}-Node{c_i{_j}}\right)^2} $
			\If {$d\left( n,Node{c_i}\right) < = r $}
			\State $n \gets Cluster\ HEAD\ Node{c_i}$
			\EndIf	
			\EndFor
			\EndFor
			\EndProcedure
			\caption{Heuristic Algorithm to Calculate Cluster and Cluster Head of FuzzyART/DBSCAN \& MEC}
			\label{alg}
		\end{algorithmic}
	\end{algorithm}

	The FuzzyART, DBSCAN and MEC AI/ML are unsupervised learning clustering techniques that separates UEs into clusters (hence implement ultra-dense networks) under the BS, by utilizing distances, like the Euclidean Distance, as a metric. Then, the heuristic algorithm, that we developed (and presented in Algorithm \ref{alg}), utilizes the clustering results extracted by these approaches, and selects the D2D Device in the identified clusters with the best Data Rate to be set as D2D Relay node and made D2D Relay Cluster Head (CH). Once the D2D Relay CH is selected, the algorithm assigns the UEs within a radius of 200m (WIFI Direct) from the D2D Relay CH, to become D2D Clients of the cluster and connect to it. Also, UEs not within the radius will stay connected to the BS (non-D2D-UEs). Our AI/ML heuristic algorithm return the clusters with CH as D2D Relays. The connected D2D Clients at the CH and the resulting non-D2D UEs.

	\section{Performance Evaluation} \label{performance_evaluation}
	This section examines, evaluates, and compares the efficiency of DAIS with the other investigated approaches, under a D2D communications network with a small number of UEs.
	
	\subsection{Methodology}\label{methodology}
	Firstly, the performance of DAIS for a scenario with a small number of D2D Devices (\textless= 200), as compared to the number of D2D Devices in \cite{Ioannou2020} which increase up to 1000, is investigated, while varying the device Battery Power Level and the WDR thresholds. For this, a “brute force” investigation of the aforesaid thresholds was executed with values from 0\% to 100\% using a step of 5\%.
	
	Secondly, the effect the Transmission Power (TP) has on the investigated approaches, in terms of overall total PC and total SE achieved, is also investigated and demonstrated. For the communication power a “brute force” investigation was executed with values from 160 mW to 60 mW using a decreasing step of 10 mW.
	
	Thirdly, the FuzzyART, DBSCAN and MEC AI/ML unsupervised learning clustering techniques are compared with the Sum Rate Approach, the DAIS algorithm, the non-D2D UE and the Random clustering approach (as shown in \cite{Ioannou2020}) in a D2D communication network in terms of SE and PC. The case where D2D communication is not used is also compared (we refer to this as non-D2D-UE approach). 

	Finally, we examine each approach according to the cluster formation quality (as shown in the \ref{clusterformation}).
	
	\subsection{Simulation Environment}
	In order to investigate how to achieve the best results in a network with a low number of D2D devices, a range of 1 to 200 D2D Devices were used. The devices are placed in a cell range of 1000 meter radius from the BS using a Poisson Point Process distribution model. In our simulation environment we keep the same comparison measurements of performance and these are the Total SE (Sum rate), Total PC and Execution Time as in \cite{Ioannou2020}. Also, the Channel State Information (CSI) used in the investigation is the Statistical CSI. In addition we keep the same formulas for D2D UEs battery power level estimation and WDR and the same simulation constraints and simulation parameters. However, we introduce new constraints and parameters as in section \ref{problem_formulation_and_approaches_specifics}. The simulation environment is implemented in Java (i.e. Java 11.0 with Apache Netbeans 11.6 IDE) using the JADE Framework\cite{JADE}, LTE/5G Toolbox libraries from Matlab (2020a) and also the SMILE library that is used for AI/ML implementation. The hardware used for the simulation is the following: i) an Intel(R) Core(TM) i7-8750H CPU @ 2.20GHz; ii) 24 GB DDR4; iii) 1TB SSD hard disk; and iv) NVIDIA GeForce GTX 1050 Ti graphics card with 4GB DDRS5 memory.

	\subsection{Results}
	
	\subsubsection{Evaluation of DAIS Approach}
	
\begin{figure}
		\centering
		\captionsetup{skip=0pt}
		\includegraphics[width=0.9\linewidth, height=0.25\textheight]{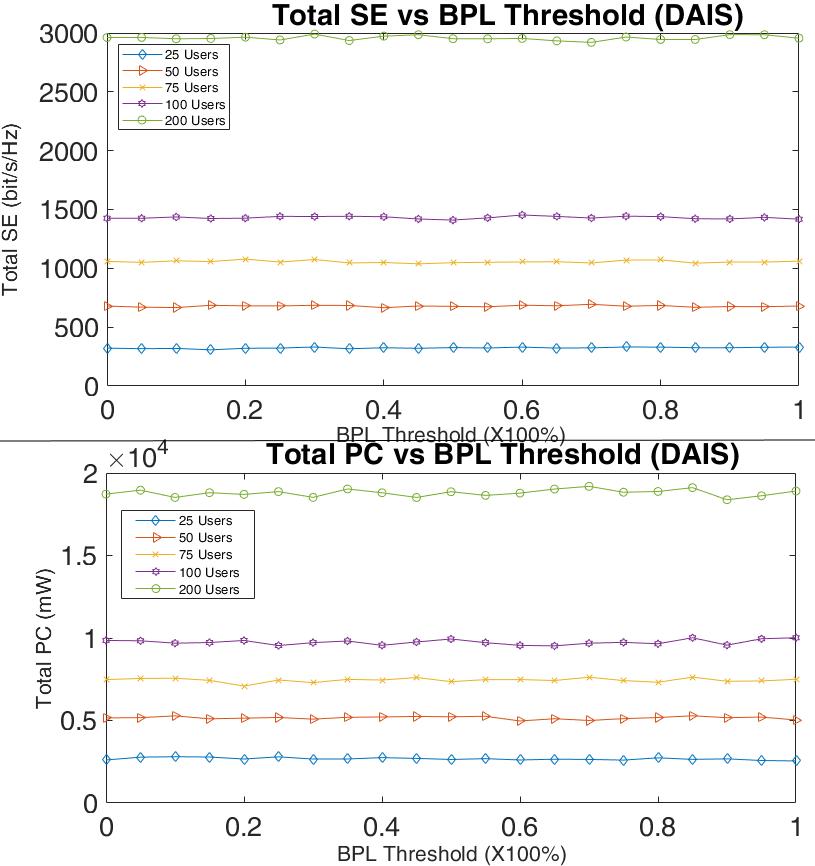}
		\caption{Total SE \& Total PC vs Battery Power Level (BPL) Threshold}
		\label{fig:totalspectralefficiencypowerneededvspowerthresholdofdifferentapproachestreportusers2}
	\end{figure}

\begin{figure}
		\centering
		\captionsetup{skip=0pt}
		\includegraphics[width=0.9\linewidth, height=0.25\textheight]{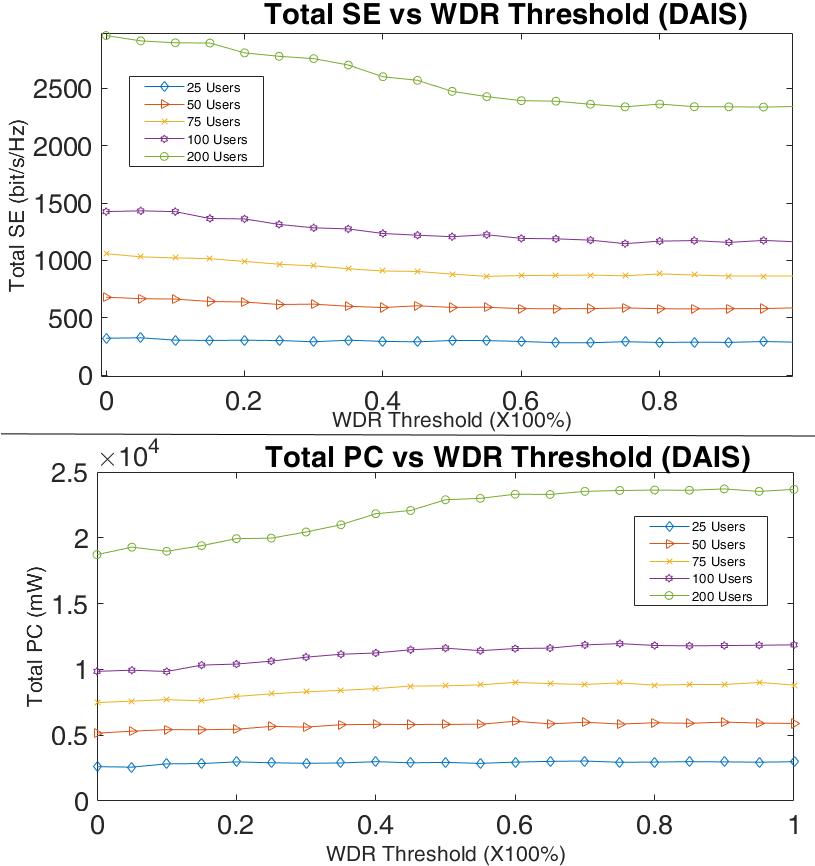}
		\caption{Total SE \& Total PC vs WDR Threshold}
		\label{fig:totalspectralefficiencypowerneededvswdrofdifferentapproachestreportusers2}
	\end{figure}

	The results related to the performance of DAIS are illustrated in Fig. \ref{fig:totalspectralefficiencypowerneededvspowerthresholdofdifferentapproachestreportusers2} and Fig. \ref{fig:totalspectralefficiencypowerneededvswdrofdifferentapproachestreportusers2}. Note that for the results provided, a “brute force” investigation was executed, by varying the Device Battery Power Level (in \%) and the Weighted Data Rate (WDR) Thresholds with values from 0\% to 100\% using a step of 5\%. During this investigation the optimum thresholds were also selected. As observed from the results, varying the Device Battery power level threshold does not cause noticeable changes on the total PC nor the sum rate (i.e., total SE).
	
	On the other hand, by varying the WDR Threshold, we observe that the results are considerably affected, in terms of SE and PC. More specifically, with a different number of D2D Devices and different values for the WDR threshold there are major changes in the resulting total PC and total SE. However, in order to achieve these results at least a number of 75 D2D Devices must exist under the BS.	Furthermore, the WDR threshold value achieving optimized results is 20\% (see section \ref{methodology} for an explanation on the use of this threshold).
	
	\subsubsection{Effect of Transmission Power Alteration on the Investigated Approaches}
	
\begin{figure*}
		\begin{subfigure}{.5\textwidth}
			\centering
			\includegraphics[width=1\linewidth]{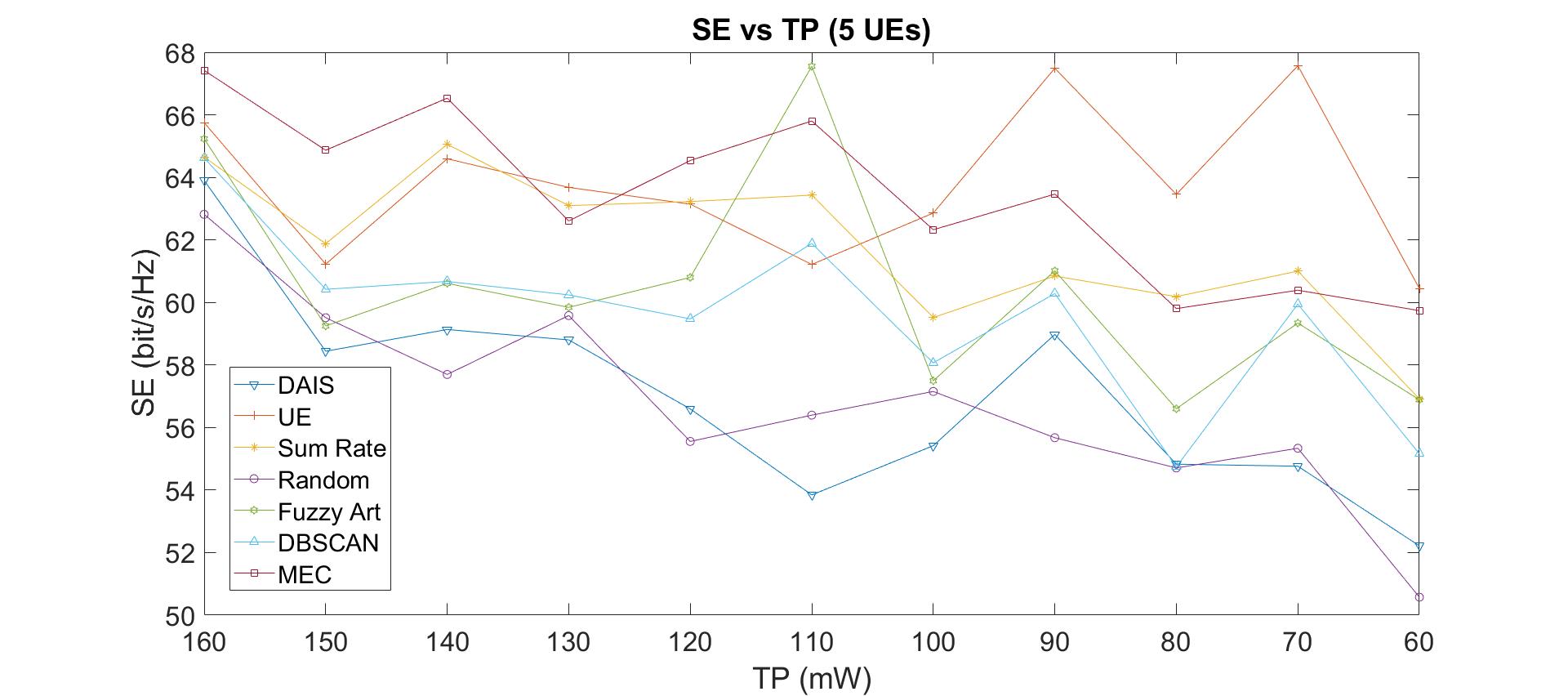}  
			\caption{SE vs TP (5 UEs)}
			\label{fig:sub-firstSETP}
		\end{subfigure}
		\begin{subfigure}{.5\textwidth}
			\centering
			\includegraphics[width=1\linewidth]{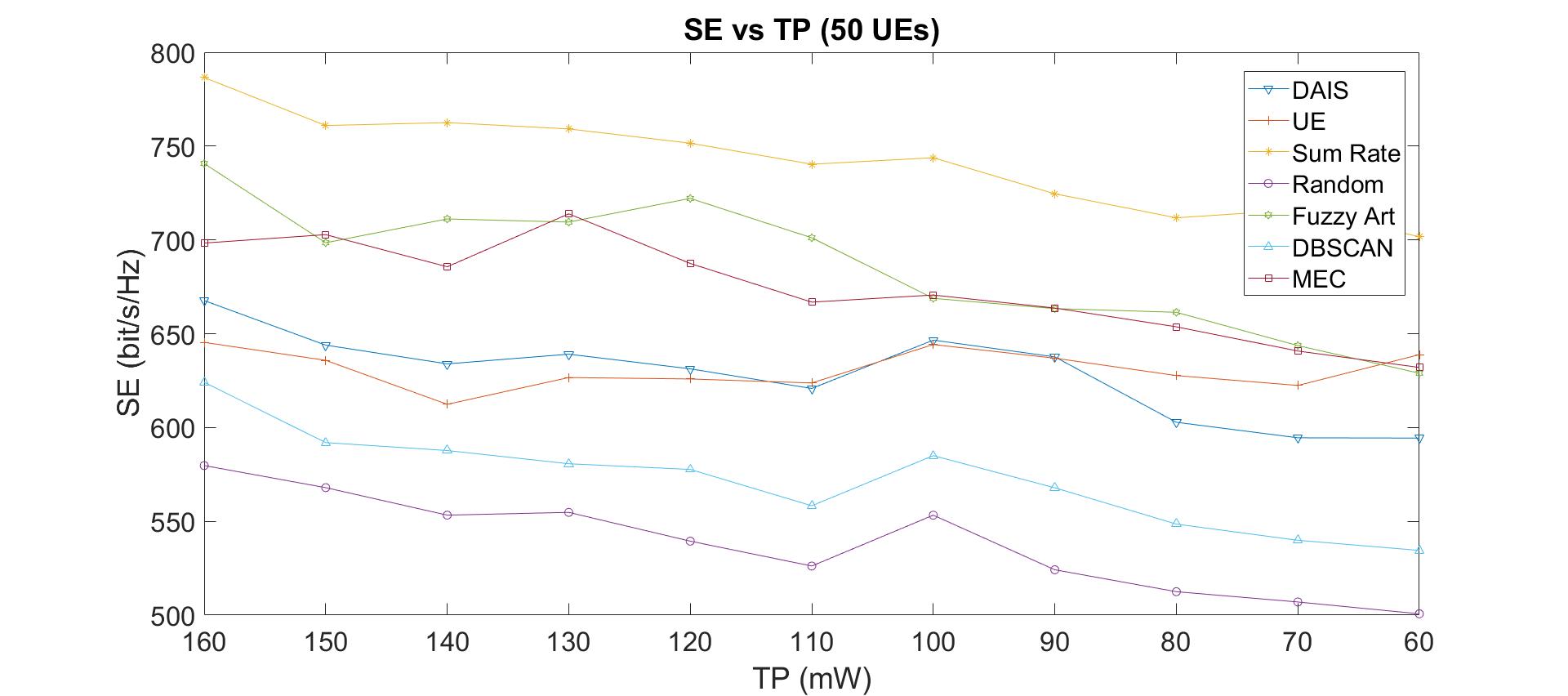}  
			\caption{SE vs TP (50 UEs)}
			\label{fig:sub-secondSETP}
		\end{subfigure}

		\begin{subfigure}{.5\textwidth}
			\centering
			\includegraphics[width=1\linewidth]{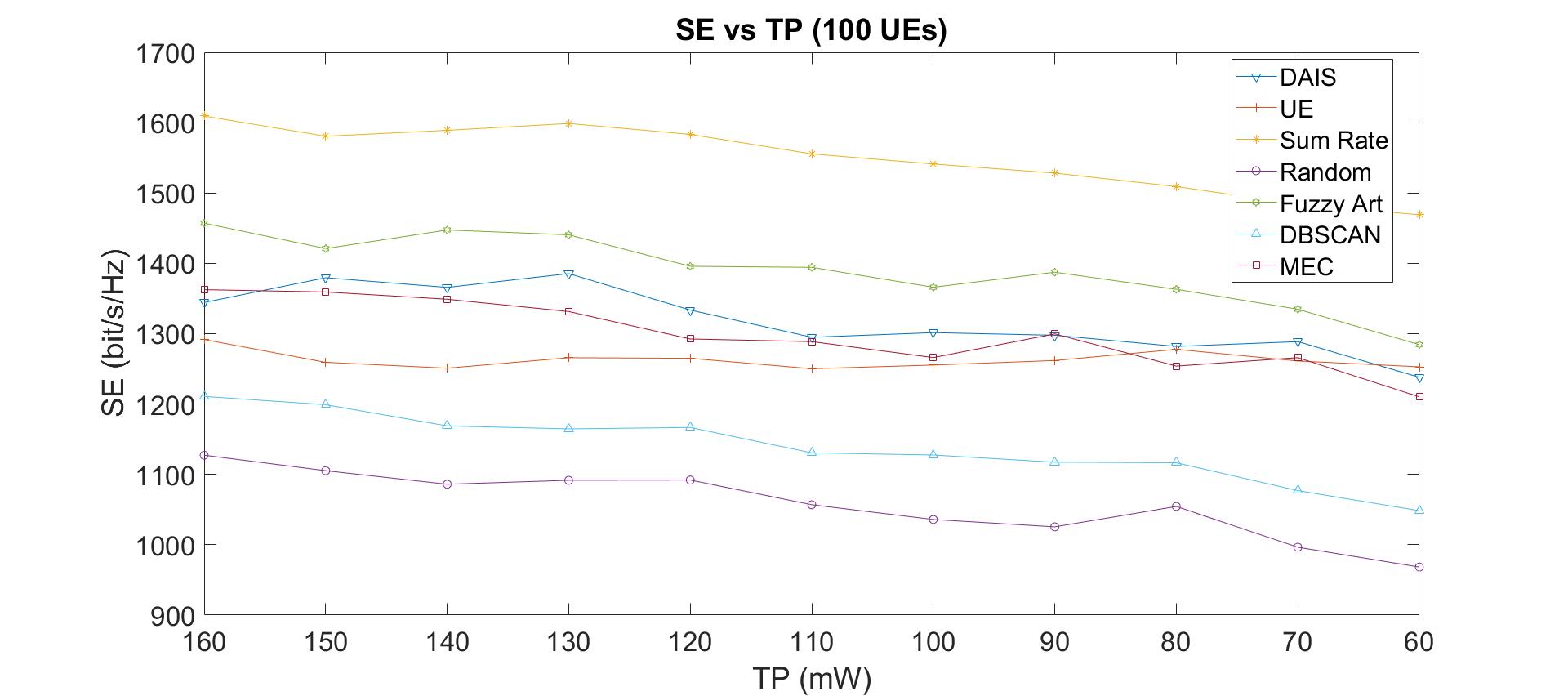}  
			\caption{SE vs TP vs SE (100 UEs)}
			\label{fig:sub-thirdSETP}
		\end{subfigure}
		\begin{subfigure}{.5\textwidth}
			\centering
			\includegraphics[width=1\linewidth]{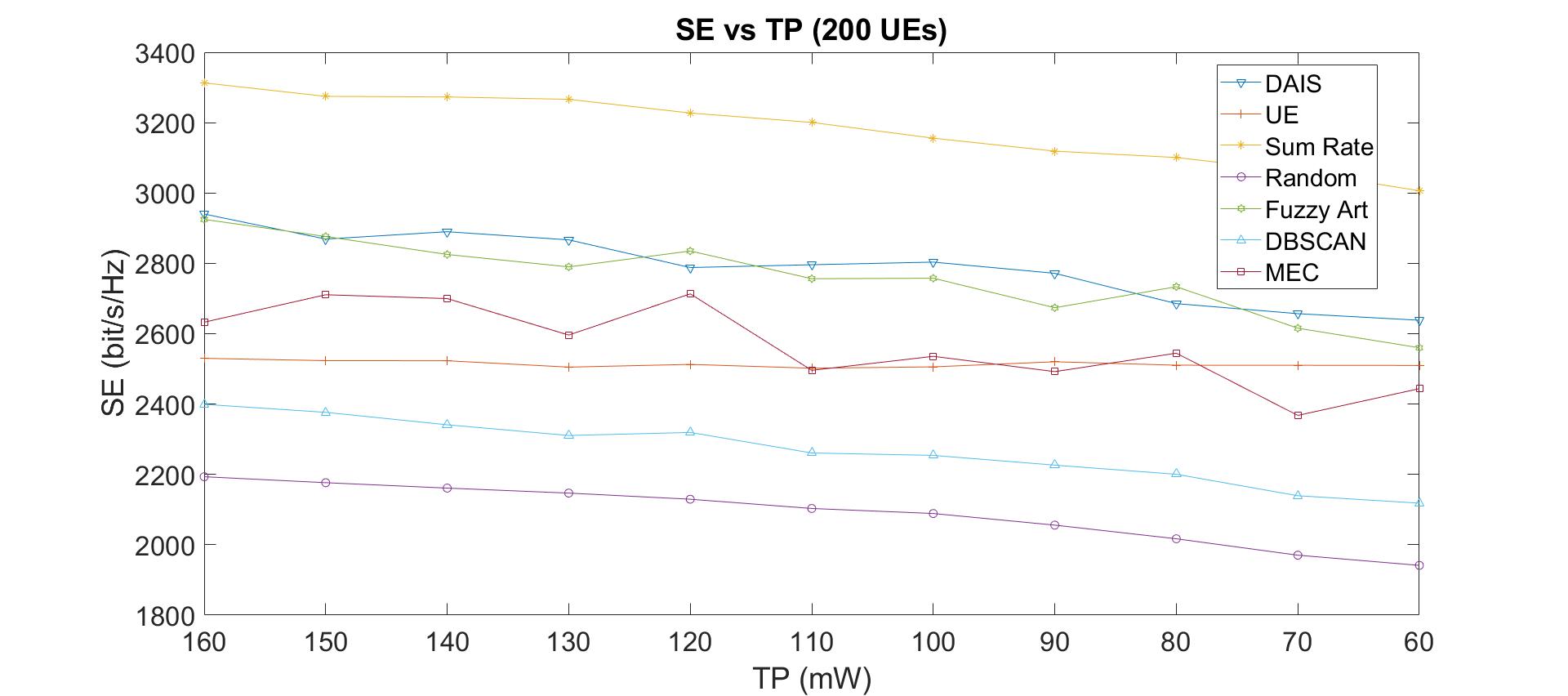}  
			\caption{SE vs TP (200 UEs)}
			\label{fig:sub-fourthSETP}
		\end{subfigure}
		\caption{Total SE vs Transmission Power (TP) of Different Approaches 5/50/100/200 UEs}
		\label{fig:totalspectralefficiencyvstransmissionpowerofdifferentapproachestreportusers}
	\end{figure*}

\begin{figure*}
		\begin{subfigure}{.5\textwidth}
			\centering
			\includegraphics[width=1\linewidth]{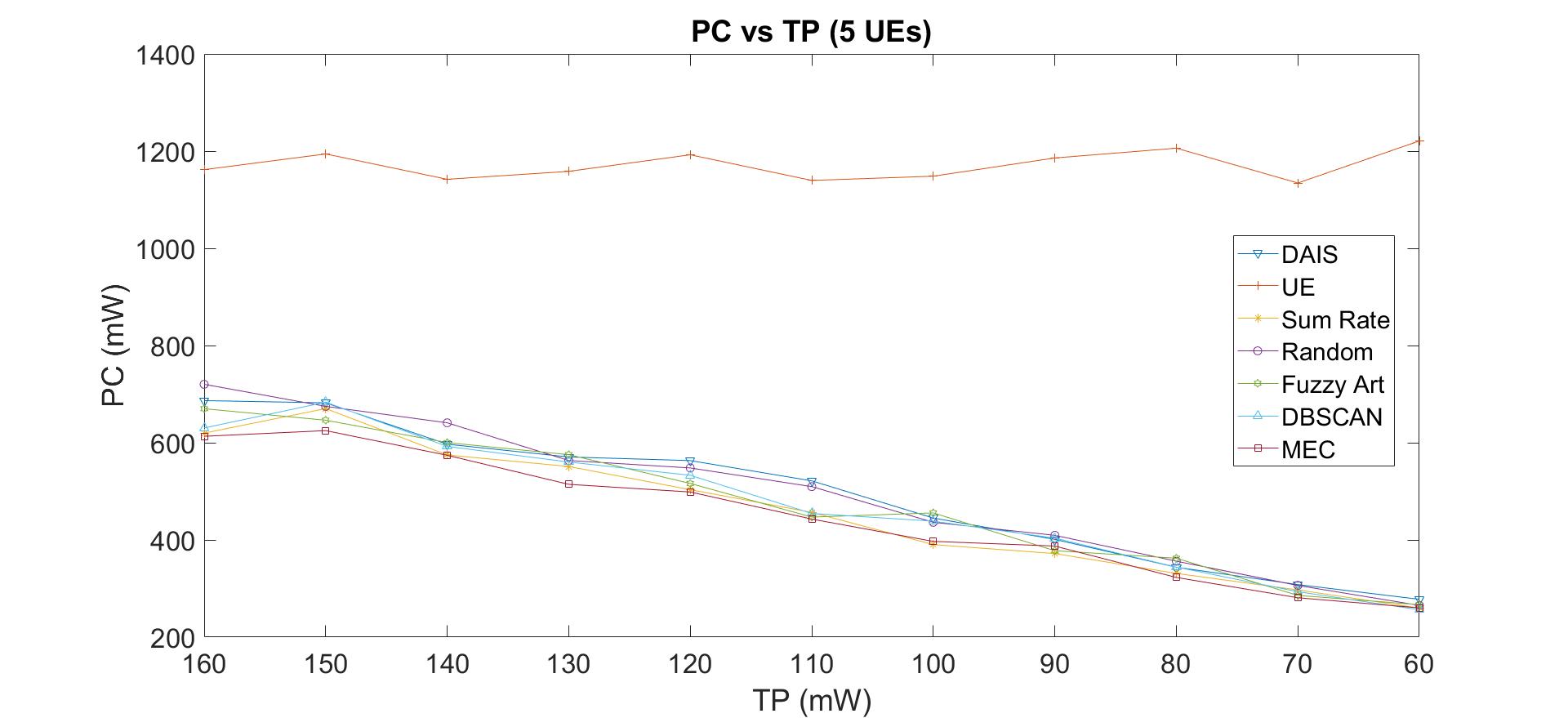}  
			\caption{PC vs TP (5 UEs)}
			\label{fig:sub-firstPCTP}
		\end{subfigure}
		\begin{subfigure}{.5\textwidth}
			\centering
			\includegraphics[width=1\linewidth]{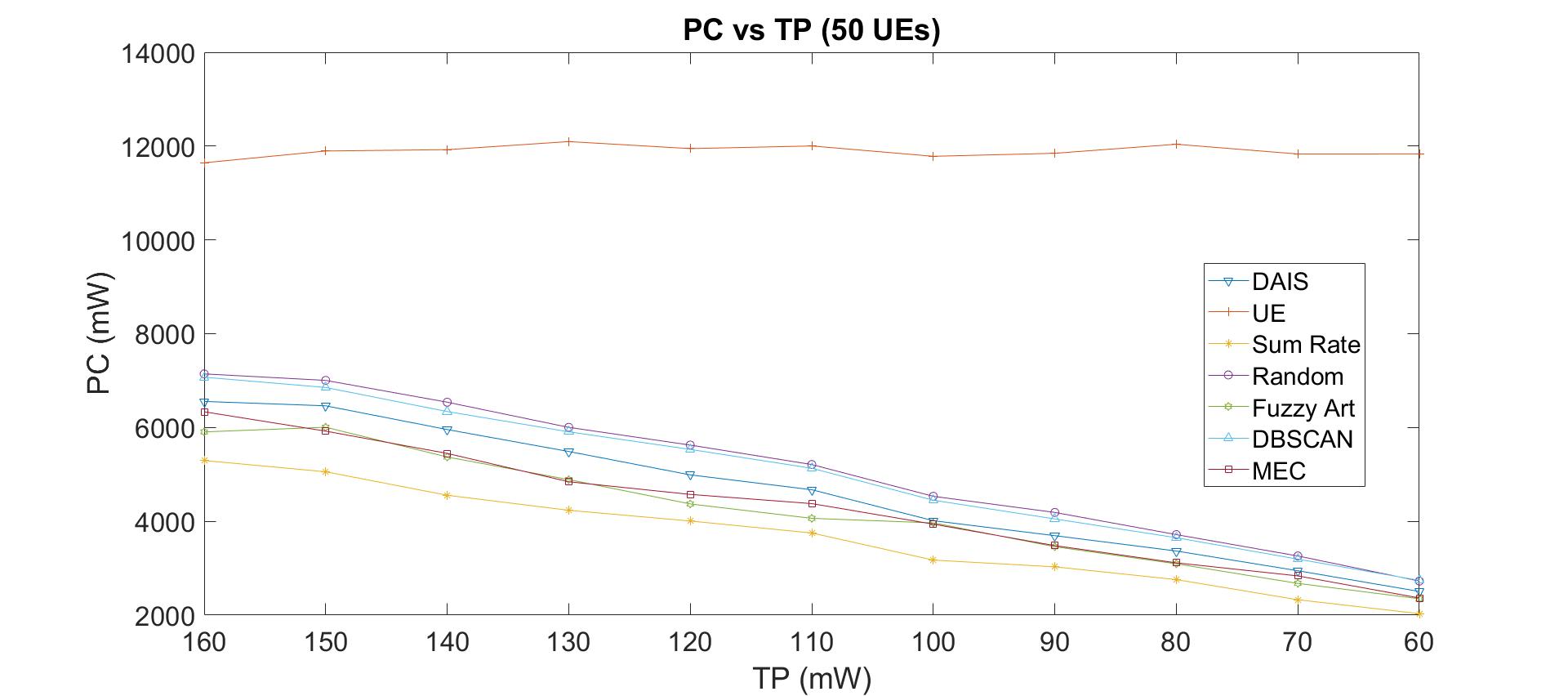}  
			\caption{PC vs TP (50 UEs)}
			\label{fig:sub-secondPCTP}
		\end{subfigure}

		\begin{subfigure}{.5\textwidth}
			\centering
			\includegraphics[width=1\linewidth]{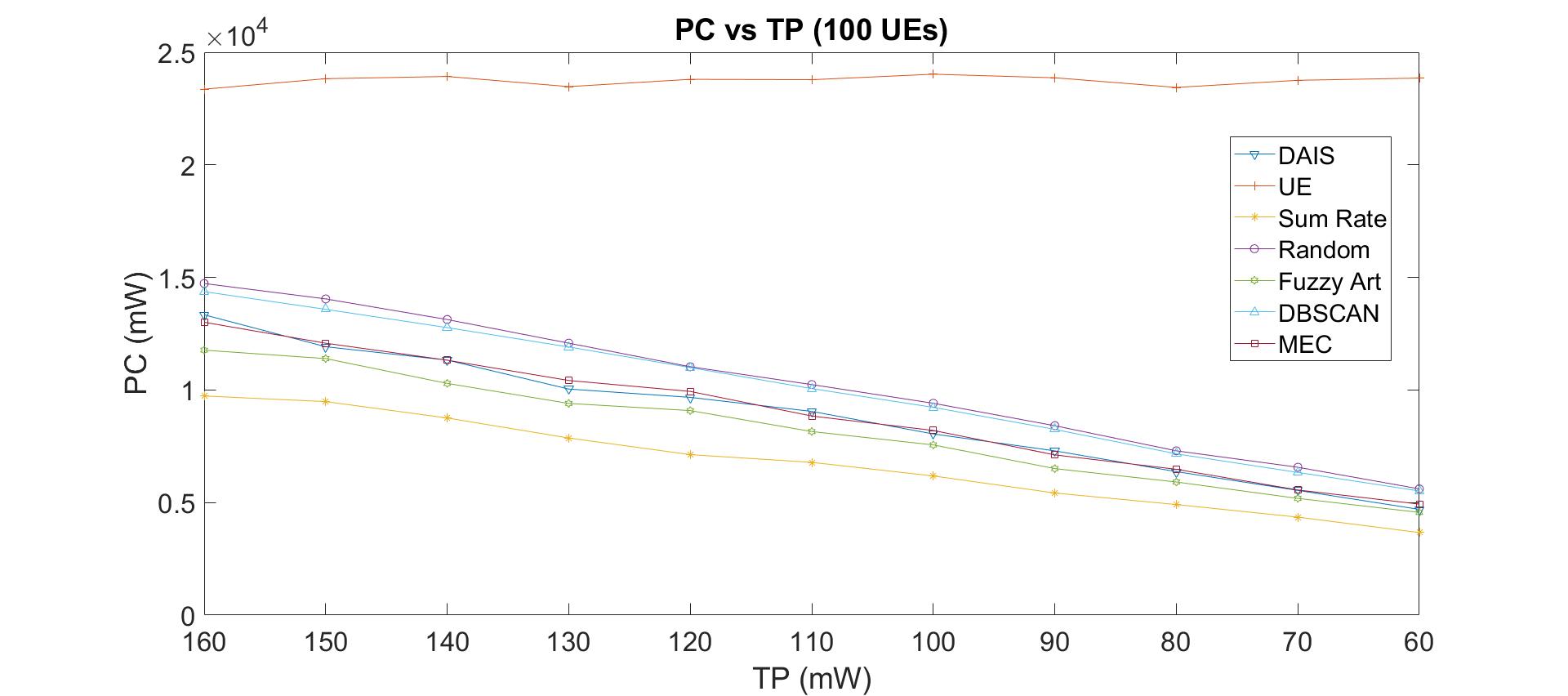}  
			\caption{PC vs TP vs SE (100 UEs)}
			\label{fig:sub-thirdPCTP}
		\end{subfigure}
		\begin{subfigure}{.5\textwidth}
			\centering
			\includegraphics[width=1\linewidth]{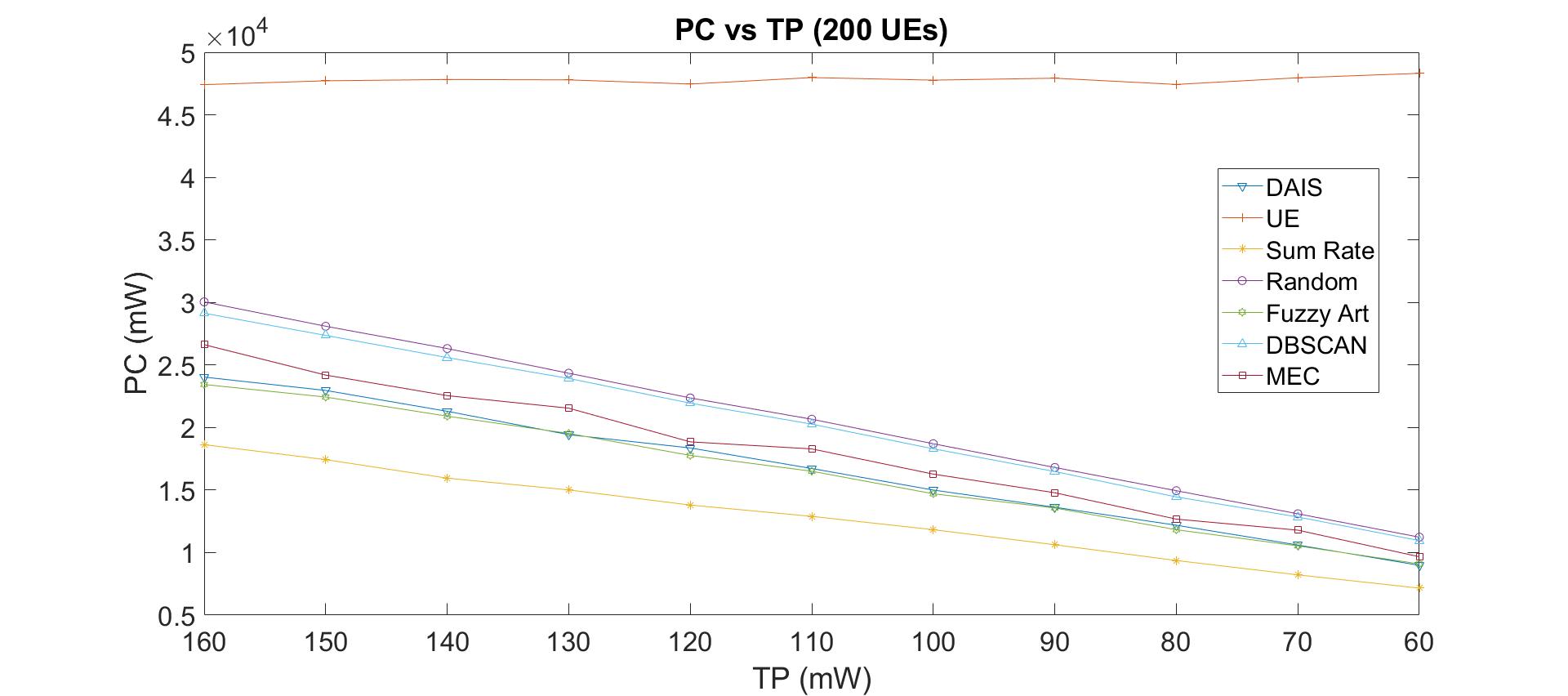}  
			\caption{PC vs TP (200 UEs)}
			\label{fig:sub-fourthPCTP}
		\end{subfigure}
		\caption{Total PC vs Transmission Power (TP) of Different Approaches 5/50/100/200 UEs}
		\label{fig:totalpowerneededvstransmissionpowerofdifferentapproachestreportusers}
	\end{figure*}

	The effect that the transmission power has on the investigated approaches, in terms of total PC and total SE (sum rate) achieved, are illustrated in Fig. \ref{fig:totalspectralefficiencyvstransmissionpowerofdifferentapproachestreportusers} and Fig. \ref{fig:totalpowerneededvstransmissionpowerofdifferentapproachestreportusers}. 
	
	As observed, by altering the transmission power of the communication and the number of UEs (D2D Devices) gains are provided on the total PC with a small trade off on the SE. More specifically, by altering (decreasing) the transmission power, the following observations are made: i) for the scenarios with low number of UEs (i.e., up to 100 UEs), there is noticeable improvement on the total network PC (i.e., up to 64.10\% for DAIS), with a small decrease on the SE (i.e., a maximum of 20\% decrease for DBSCAN); ii) for the scenarios with more than 100 UEs, significant gains are also observed on the total PC (i.e., up to 66.10 \% decrease for MEC) but with minor decrease on the SE (i.e., a maximum of 13\% decrease of Random). In addition, for all approaches compared (except the non-D2D-UE), the values of total PC change rapidly from 0 UEs to 200 UEs, but they do not have a large scale of difference in each approach. On the other hand, for the non-D2D-UE approach the total PC used compared to all other approaches is significant. The reason is that with this approach, all the UEs have direct connections with the BS, which are power consuming.

	\subsubsection{Performance Comparison of the Investigated Approaches}
	
	In this section, the performance of the approaches is compared in terms of total SE (Sum Rate) and total PC achieved. For this comparison a predefined transmission power of 160 mW is used for all approaches. The following examinations are concluded from the Table \ref{se_pc} for SE and PC.

	\begin{table}
		\centering
		\caption{Total SE and Total PC of each Approach}
		\label{se_pc}
		\resizebox{\linewidth}{!}{%
		\begin{tabular}{|l|c|c|c|c|c|c|c|} 
			\hline
			& \multicolumn{7}{c|}{\textbf{Total SE}} \\ 
			\hline
			\textbf{\# Ues} & \textbf{DAIS} & \textbf{non-D2D UE} & \textbf{Sum Rate} & \textbf{DR} & \textbf{Fuzzy ART} & \textbf{DBSCAN} & \textbf{MEC} \\ 
			\hline
			5 & 63.90 & \textbf{65.75} & 64.65 & 62.82 & 65.23 & 64.62 & \textbf{67.42} \\ 
			\hline
			50 & 667.67 & 645.38 & \textbf{786.54} & 579.73 & \textbf{740.70} & 624.11 & \textbf{698.33} \\ 
			\hline
			100 & \textbf{1344.39} & 1291.69 & \textbf{1609.46} & 1127.30 & \textbf{1457.10} & 1210.86 & \textbf{1362.60} \\ 
			\hline
			200 & \textbf{2940.20} & 2529.99 & \textbf{3312.99} & 2193.42 & \textbf{2924.93} & 2399.22 & 2632.69 \\ 
			\hline
			& \multicolumn{7}{c|}{\textbf{Total PC}} \\ 
			\hline
			\textbf{\# Ues} & \textbf{DAIS} & \textbf{non-D2D UE} & \textbf{Sum Rate} & \textbf{DR} & \textbf{Fuzzy ART} & \textbf{DBSCAN} & \textbf{MEC} \\ 
			\hline
			5 & 686.66 & 1161.88 & \textbf{619.83} & 720.22 & 670.02 & 630.39 & \textbf{613.24} \\ 
			\hline
			50 & 6555.84 & 11644.01 & \textbf{5295.86} & 7143.10 & \textbf{5906.62} & 7072.72 & \textbf{6335.16} \\ 
			\hline
			100 & 13340.20 & 23361.62 & \textbf{9734.12} & 14729.55 & \textbf{11769.16} & 14367.71 & \textbf{13008.63} \\ 
			\hline
			200 & \textbf{24030.83} & 47412.57 & \textbf{18627.00} & 30046.24 & \textbf{23439.71} & 29136.61 & 26622.97 \\
			\hline
		\end{tabular}
		}
	\end{table}

	Starting from Sum Rate Approach, in terms of the total power needed (i.e., power consumption), the best results are provided by it, while the worst performance is observed for the non-D2D-UE approach. Also, all approaches are relatively close, in terms of total SE from a range of UEs of 0 to 50. Beyond 50 UEs, the DAIS and Sum Rate Approach, approaches start to show increased SE and they conclude to have better SE than other centralized AI approaches. Therefore, through Sum Rate Approach we show that it is valuable to implement back-hauling and include D2D Multi hop Relay to be selected as mode when we have more than 50 UEs under BS (exception is when devices exists out of BS range/coverage than a D2D MHR is needed). This is shown in the Table \ref{table:d2dmhr}, where Sum Rate Approach results to more than 1 D2DMHR after 50 UEs. 
	
	In terms of SE, DAIS seems to under-perform compared to the other approaches for a network with a small number of devices (i.e., 10 UEs). However, from 50 UEs and above, DAIS is better than the DBSCAN, Random, and non-D2D-UE approaches. Finally, at 200 UEs (maximum number of UEs examined) DAIS shows its benefits by reaching the results provided by the Sum Rate Approach. Continuing our examination on total PC, DAIS outperforms the non-D2D-UE approach for all number of devices examined. Furthermore, at 200 devices the DAIS is better than DBSCAN, MEC, non-D2D-UE approach, and Random, but it has the same total PC with FuzzyART. Also, the Table \ref{table:d2dmhr} shows that even that DAIS results to 8 D2DMHR at 50 UEs it is not achieving the best results in terms of SE and PC. 

	The MEC is performing the best results in terms of SE and PC until 10 UEs, from 10 UEs until 100 UEs is achieving better results over all other approaches except FuzzyART and Sum Rate approaches. Also, for more than 100 UEs is better than DBSCAN, Random, and non-D2D UE.

	The FuzzyART is performing the best results in terms of SE and PC from 10 UEs until 100 UEs is achieving better results over all other approaches except the Sum Rate approach. Additionally, for more than 100 UEs is better than all other approaches except DAIS, that have similar results and expect Sum Rate Approach that comes first.	   

	The DBSCAN is performing better results in terms of SE than Random for all UEs. In terms of PC, it is performing better results than Random and non-D2D UE for all UEs.		
	
	The non-D2D-UE approach has the worst performance in terms of total PC, compared to all other related approaches. Nevertheless, in terms of SE, it provides the better results compared to DAIS and DBSCAN when the number of UEs in the Network is 10 or less (i.e., 5 UEs). This the reason that this investigation sets the limit of achieving better results for a clustering approach the 5 UEs. Additionally, non-D2D UE for more UEs provides the worst results in terms of SE except the Random. Random approach is always the worst than all other approaches in terms of SE. However, Random provides better performance in terms of total PC compared to the non-D2D-UE approach but worst over others.	
	
	In our analysis, we examine the mean time of execution of each approach (centralized, distributed, semi-distributed, and DAI) per Device. The faster approach is the DAIS (DAI) with 100 ms with any UE (from 1..200 UEs), the second faster approach is the DR, then the MEC, DBSCAN (centralized), and Sum Rate Approach (distributed) are following as shown in the Table \ref{table:meantime}.
	
	\begin{table}
	\centering
	\caption{Time of execution of each approach (1 = 100 ms)}
	\label{table:meantime}
	\resizebox{\linewidth}{!}{%
	\begin{tabular}{lcccccc} 
		\toprule
		\begin{tabular}[c]{@{}l@{}}\textbf{Number }\\\textbf{of }\\\textbf{Devices}\end{tabular} & \textbf{DAIS} & \begin{tabular}[c]{@{}c@{}}\textbf{Sum Rate }\\\textbf{Approach}\end{tabular} & \textbf{DR} & \textbf{FuzzyART} & \textbf{DBSCAN} & \textbf{MEC} \\
		50 & 0 & 1 & 0 & 0 & 1 & 15 \\
		100 & 1 & 12 & 0 & 2 & 7 & 36 \\
		200 & 1 & 12 & 0 & 2 & 7 & 36 \\
		\bottomrule
	\end{tabular}
	}
	\end{table}

    \begin{table}
    \centering
	\caption{Resulting D2DMHR and D2DMHR without Devices to Share}
	\label{table:d2dmhr}
        \begin{tabular}{|l|c|c|c|} 
        \hline
        \textbf{D2D Devices} & \multicolumn{3}{c|}{\textbf{D2D MHR}} \\ 
        \hline
         & \textbf{DAIS} & \textbf{Sum Rate} & \textbf{Random} \\ 
        \hline
        5 & 0 & 0 & 0 \\ 
        \hline
        50 & 8 & 1 & 10 \\ 
        \hline
        100 & 13 & 2 & 22 \\ 
        \hline
        200 & 17 & 1 & 41 \\ 
        \hline
        \multicolumn{1}{|c|}{\textbf{D2D Devices}} & \multicolumn{3}{c|}{\textbf{D2DMHR with No Sharing Devices}} \\ 
        \hline
         & \textbf{DAIS} & \textbf{Sum Rate} & \textbf{Random} \\ 
        \hline
        5 & 5 & 3 & 2 \\ 
        \hline
        50 & 15 & 12 & 6 \\ 
        \hline
        100 & 19 & 7 & 11 \\ 
        \hline
        200 & 24 & 12 & 22 \\
        \hline
        \end{tabular}
    
    \end{table}

	\subsubsection{Examination of Clusters Formation Quality}
	\label{clusterformation}
	Additionally, in our examination we investigated some extra characteristics of each algorithm and compared the performance of the different approaches in terms of number of messages exchanged in order one approach to conclude, number of resulting non-D2D UEs, number of clusters formed and total number of devices under cluster. The results are provided in Table \ref{Table:numberofClusters}. 

	Regarding the number of messages that each approach needs to exchange in order to conclude on the Transmission mode selection for all runs\footnote{Run is the execution of the algorithm with a different number of UEs in each instance of the scenario}, from the worst to best performance is provided by Sum Rate Approach, FuzzyART, MEC, DBSCAN, DAIS and Random. 

	Additionally, by examining the resulting number of non-D2D UEs for all runs, at DAIS, Sum Rate Approach and Random approach all devices enter the D2D network and become D2D Devices. For the rest of the approaches, FuzzyART has the least number of resulting non-D2D UEs followed by MEC and DBSCAN.

	In terms of the created clusters, the total number of users that are served by cluster \footnote {D2D Relay/D2D Multi Hop Relay that are directly connected to BS are not included.} and number of clusters created per approach are investigated.

 	Therefore, by investigating the clusters density (mean number of devices per cluster) in incremental order, the following results are provided: i) for 50 UEs, DBSCAN comes first and then MEC if following; ii) for 100 UEs, DAIS comes first and then by DBSCAN; and iii) for 200 UEs DAIS come first and then DBSCAN. Additionally, by investigating the number of clusters in incremental order for 50-200 UEs, DAIS comes first and then FuzzyART is following with Sum Rate Approach to be next. 

	\begin{table}
		\centering
		\caption{Cluster Quality}
		\label{Table:numberofClusters}
		\resizebox{\linewidth}{!}{%
		\begin{tabular}{|l|c|c|c|c|c|c|} 
			\hline
			& \textbf{ DAIS } & \begin{tabular}[c]{@{}c@{}}\textbf{ Sum Rate }\\\textbf{Approach }\end{tabular} & \textbf{ FuzzyART } & \textbf{MEC} & \begin{tabular}[c]{@{}c@{}}\textbf{Distributed}\\\textbf{Random }\end{tabular} & \textbf{ DBSCAN } \\ 
			\hline
			\# of Devices & \multicolumn{6}{c|}{\textbf{\# Messages }} \\ 
			\hline
			\textbf{50} & \textbf{65 } & 1336 & 143 & 104 & \textbf{0 } & \textbf{74 } \\ 
			\hline
			\textbf{100} & \textbf{121 } & 5165 & 294 & 213 & \textbf{1 } & \textbf{151 } \\ 
			\hline
			\textbf{200} & \textbf{230 } & 20321 & 595 & 414 & \textbf{2 } & 300 \\ 
			\hline
			\# of Devices & \multicolumn{6}{c|}{\textbf{\# non-D2D UEs }} \\ 
			\hline
			\textbf{50} & \textbf{0 } & \textbf{0 } & 14 & 22 & \textbf{0 } & 38 \\ 
			\hline
			\textbf{100} & \textbf{0 } & \textbf{0 } & 34 & 42 & \textbf{0 } & 75 \\ 
			\hline
			\textbf{200} & \textbf{0 } & \textbf{0 } & 43 & 92 & \textbf{0 } & 150 \\
			\hline
			\multicolumn{1}{|c|}{\# of Devices} & \multicolumn{6}{c|}{\textbf{\textbf{Total \# Cluster Devices}}} \\ 
			\hline
			\textbf{50} & 6 & 6 & 6 & \textbf{9 } & 0 & \textbf{10 } \\ 
			\hline
			\textbf{100} & \textbf{97 } & 13 & 17 & 17 & 4 & \textbf{25 } \\ 
			\hline
			\textbf{200} & \textbf{146 } & 26 & 34 & 38 & 21 & \textbf{49 } \\ 
			\hline
			\# of Devices & \multicolumn{6}{c|}{\textbf{\# Clusters }} \\ 
			\hline
			\textbf{50} & \textbf{13 } & \textbf{12 } & 8 & 5 & 1 & 1 \\ 
			\hline
			\textbf{100} & \textbf{19 } & \textbf{17 } & 7 & 5 & 3 & 1 \\ 
			\hline
			\textbf{200} & \textbf{26 } & \textbf{25 } & 7 & 4 & 7 & 1 \\ 
			\hline
		\end{tabular}
		}
	\end{table}

	Overall results in Clusters Quality shows that due to the low number of UEs, even if we have good characteristic values of each approach in terms of cluster Quality the approaches that are achieving good results they do not achieve good results in terms of Total SE/PC. The reason is because the number of UEs under the mobile network is small.

	\subsubsection{Overall Remarks}
	\label{overall}
	The research objective on this paper is threefold. Firstly, it examines the performance of the DAIS algorithm with the proposed changes in threshold (i.e., WDR Threshold), in terms of SE and PC, considering scenarios with a small number of Devices (i.e., \textless= 200). During this examination, the WDR and the BPL DAIS' thresholds, affecting the SE and PC of the network, have been examined and values achieving best performance have been determined. Secondly, it introduces the use of unsupervised learning AI/ML approaches in Transmission mode selection in D2D Communication and compares the performance of DAIS with FuzzyART, DBSCAN and MEC as well as other related approaches (i.e., Distributed Random, Distributed Sum Rate Approach, Centralized non-D2D-UE). Last, it examines the effect the transmission power has on the investigated approaches, in terms of PC and SE achieved. 	
	
	The results obtained demonstrated that DAIS, compared to all other related approaches, with the right tuning of WDR and BPL threshold values, can provide significant gains in terms of SE, PC, and  time of execution. More precisely, the results showed that DAIS and Sum Rate Approach outperformed all other approaches in terms of SE. FuzzyART, DAIS and Sum Rate Approach outperformed all other related approaches in terms of PC. Additionally, our findings showed that, by reducing the transmission power of communication for all approaches, the SE and PC of the network is significantly affected (SE in a negative way and PC in a positive way) when the amount of UEs is less than 100. On the other hand, from 100 to 200 UEs the effect on SE becomes smoother while on PC the gains remain the same. 
		
	In the performance comparison provided above the different investigated approaches are evaluated in terms of SE and PC. The results illustrated that the worst performance is provided by the Random approach, while the best performance is provided by Sum Rate Approach, FuzzyART and DAIS. On the other hand, in terms of total PC, the worst performance is provided by non-D2D-UE approach, while best is provided again by the Sum Rate Approach, DAIS and FuzzyART. Additionally, the paper shows that unsupervised learning approaches such as FuzzyART can be used for transmission mode selection in D2D Communication. Additionally, we show that DBSCAN is not achieving good results in less than 200 UEs in terms of SE compared to non-D2D UE. However, it achieves great results in terms of PC compared to non-D2D UE.
	
	Therefore, some of the approaches are good alternatives to be used for Transmission mode selection in the D2D communication the faster approach is the DAIS over all other and then DR  follows along with non-D2D UE, FuzzyART, DBSCAN, Sum Rate and then MEC. In addition, the following findings extracted from Table \ref{se_pc}: i) Some of the 5G requirements are achievable through Transmission Mode Selection (i.e. High Data Rates, Low Power Consumption); ii) The critical point that SE, PC gains increases rapidly is 100 UEs for all approaches; iii) coverage expansion is achieved; and iv) the lower limit of all approaches is 5 UEs in order to form clusters and 50 UEs in order to form back-hauling network.
	
	\section{Conclusions and Future Work}\label{conclussions_and_future_work}
	
	Overall, results demonstrate the exceptional performance of DAIS, compared to all other related approaches in terms of SE, PC, execution time and cluster formation. Also, results show that the investigated AI/ML approaches are also beneficial for Transmission Mode Selection (TMS) in 5G D2D communication, even with a smaller number of devices (i.e., \textgreater=5 for clustering, \textgreater=50 for back-hauling) as a lower limits. A full examination of the results is shown at \ref{overall}. As future work we will investigate the performance of the same AI/ML approaches in scenarios with large number of UEs (i.e., up to 1000 UEs under the same BS) considering non ideal CSI in D2D communication network. 
	
	\bibliography{conference_ieee}
	\bibliographystyle{ieeetr}
	
\end{document}